\documentclass[10pt,conference]{IEEEtran}

\usepackage{amsmath,amssymb,eucal,graphicx}
\usepackage{epsfig}
\usepackage{exscale}
\usepackage{latexsym}
\usepackage{verbatim}
\usepackage{amstext}
\usepackage{latexsym}
\usepackage{color}
\usepackage{ifthen}
\usepackage{multirow}
\usepackage{subfigure}

\newtheorem{thm}{Theorem}
\newtheorem{lemma}{Lemma}

\newtheorem{defn}{Definition}[section]

\newcommand{\beq}{\begin{equation}}
\newcommand{\eeq}{\end{equation}}

\newcommand{\lp}{ \left(}
\newcommand{\rp}{ \right)}

\long\def\comment#1{}

\newfont{\bbb}{msbm10 scaled 700}

\newfont{\bb}{msbm10 scaled 1100}

\def\BibTeX{{\rm B\kern-.05em{\sc i\kern-.025em b}\kern-.08em
    T\kern-.1667em\lower.7ex\hbox{E}\kern-.125emX}}

\begin{document}

\title{Capacity region of the deterministic multi-pair bi-directional relay network}
\author{Salman Avestimehr\\
Caltech,\\
 Pasadena, CA, USA.\\
{\sffamily avestime@caltech.edu} \and
Amin Khajehnejad\\
Caltech,\\
 Pasadena, CA, USA.\\
{\sffamily amin@caltech.edu} \and
Aydin Sezgin\\
UC Irvine,\\
Irvine, CA, USA.\\
{\sffamily asezgin@uci.edu} \and
Babak Hassibi\\
Caltech,\\
 Pasadena, CA, USA.\\
{\sffamily hassibi@caltech.edu}
}


\date{\today}
\maketitle

\begin{abstract} In this paper we study the capacity region of the multi-pair bidirectional (or two-way) wireless relay network, in which a relay node facilitates the communication between multiple pairs of users. This network is a generalization of the well known bidirectional relay channel, where we have only one pair of users. We examine this problem in the context of the deterministic channel interaction model, which eliminates the channel noise and allows us to focus on the interaction between signals. We characterize the capacity region of this network when the relay is operating at either full-duplex mode or half-duplex mode (with non adaptive listen-transmit scheduling). In both cases we show that the cut-set upper bound is tight and, quite interestingly, the capacity region is achieved by a simple equation-forwarding strategy.
\end{abstract}

\section{Introduction}

Cooperative communication and relaying is one of the important research topics in wireless network information theory. The basic model to study this problem is the 3-node relay channel which was first introduced in 1971 by van der Meulen \cite{Meulen} and the most general strategies for this network were developed by Cover and El Gamal \cite{CoverGamal}. While the main focus so far is on the one-way-relay channel, bidirectional communication has also attracted attention. Bidirectional or two-way communication between two nodes was first studied by Shannon himself in~\cite{ShannonInt}. Nowadays the two-way communication where an additional node acting as a relay is supporting the exchange of information between the two nodes (or one pair) is gaining increased attention. Some relaying strategies for this one-pair two-way relay channel, such as decode-and-forward, compress-and-forward and amplify-and-forward, have been analyzed in~\cite{RankovISIT}. Network coding type techniques have been proposed by~\cite{KattiI,HauslHagenauer,BaikChung,NaryananLatticeBiDi} (and others) in order to improve the transmission rate.

The two way relay channel problem can be generalized to a multi-pair (or multiuser) setting in which the relay facilitates the communication between multiple pairs of users. In \cite{RankovMU} authors analyzed the case that the relay orthogonalizes different two-way transmissions by a distributed zero forcing algorithm and then multiple pairs communicate with each other via several orthogonalize-and-forward relay terminals. In \cite{ChenYener,ChenYenerCISS} authors investigated this problem for interference limited systems in which each pair of users share a common spreading signature to distinguish themselves from the other pairs, and proposed a jointly demodulate-and-XOR forward strategy. However, so far no attempt has been done to characterize the capacity region of this network, and the optimal relaying strategy is unknown.

In this paper we study the information theoretic capacity of the multi-pair bidirectional wireless relay network. We examine this problem in the context of the deterministic channel interaction model. The deterministic model, studied by Avestimehr, Diggavi, and Tse \cite{ADT072}, simpliÞes the wireless network interaction model by eliminating the noise and allows us to focus on the interaction between signals. This approach was successfully applied to the relay network in \cite{ADT072}, and resulted in insight in terms of transmission techniques which further led to an approximate characterization of the noisy wireless relay network problem \cite{ADTISIT08}. This approach has also been recently applied to the bidirectional relay channel problem \cite{AvestimehrSezgin}, which again resulted in finding near optimal relaying strategies as well as approximating the capacity region of the noisy (Gaussian) bidirectional relay channel.

Inspired by these results, in this paper we apply the deterministic model to the multi-pair bidirectional relay network. and analyze its capacity when the relay is operating at either full-duplex mode or half-duplex mode (with non adaptive listen-transmit scheduling). In both cases we exactly characterize the capacity region and show that the cut-set upper bound is tight. Quite surprisingly, we show that the capacity region is achieved by a simple \emph{equation-forwarding} strategy, in which different pairs are orthogonalized on the signal level space and the relay just re-orders the received equations created from the superposition of the transmitted signals on the wireless medium and forwards them.

The paper is organized as follows. In Section \ref{sec:sysModel} we state the precise definition of the problem. We will present our main results in Section \ref{sec:Main} and characterize the exact capacity region of the full-duplex and half-duplex multi-pair deterministic bidirectional relay networks. Then in Section \ref{sec:Remark} we illustrate further structures in the optimal relaying strategy  and finally conclude in Section \ref{sec:Conc}.

\section{System model \label{sec:sysModel}}
\begin{figure}
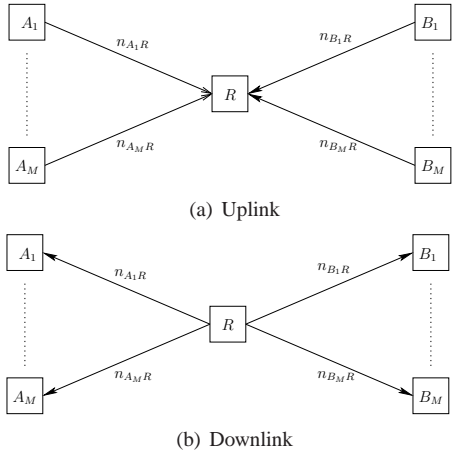

     \centering
     \subfigure[Uplink]{
  \scalebox{0.5}{   \input{sysModela.pstex_t}}
}
\subfigure[Downlink]{
       \scalebox{0.5}{\input{sysModelb.pstex_t}}
}
     \caption{The system model for $M$ pair bidirectional deterministic relay network. \label{fig:sysModel}}
\end{figure}

The system model for the M-pair bidirectional relay network is shown in Figure \ref{fig:sysModel}. In this system $M$ pairs  $(A_1,B_1),\ldots,(A_M,B_M)$ aim to use the relay to communicate with each other (\emph{i.e.} $A_1$ and $B_1$ want to communicate with each other, and so on). The relay can operate on either full-duplex or half-duplex mode. In the full-duplex mode it is able to listen and transmit at the same time, while in the half-duplex mode it can only listen or transmit at a particular time. In the half-duplex scenario, we only consider the case that the listen-transmit scheduling is non-adaptive and the relay listens a fixed $t$ fraction of the time and transmits the rest. Although $t$ can not change adaptively as a function of the channel gains, one can optimize over $t$ beforehand.

We use the deterministic channel model to model the interaction between the transmitted signals. The deterministic channel model was introduced in~\cite{ADT072}. Here is a formal definition of this channel model.

\begin{defn} \textbf{(Definition of the deterministic model)}
Consider a wireless network as a set of nodes $V$, where $|V|=N$.
Communication from node $i$ to node $j$ has a non-negative integer
gain\footnote{Some channels may have zero gain.} $n_{(i,j)}$ associated
with it. This number models the channel gain in a corresponding Gaussian setting.
At each time $t$, node $i$ transmits a vector ${\mathbf{x}_i}[t] \in
\mathbb{F}_{2}^q$ and receives a vector ${\mathbf{y}_i}[t] \in \mathbb{F}_{2}^q$ where
$q=\max_{i,j}(n_{(i,j)})$. The received signal at each node is
a deterministic function of the transmitted signals at the other
nodes, with the following input-output relation: if the nodes in the
network transmit ${\mathbf{x}_1}[t], {\mathbf{x}_2}[t] , \ldots {\mathbf{x}_N}[t]$ then the received
signal at node j, $1 \leq j \leq N$ is:
\begin{equation}
\label{eq:channel_model}
{\mathbf{y}_{j}}[t]=\sum_{k=1}^N
 {\mathbf{S}^{q-n_{k,j}}}{\mathbf{x}_{k}}[t]
\end{equation}
for all $1 \leq k \leq N$,
where $\mathbf{S}$ is the $q \times q$ shift matrix and the summation and
multiplication is in $\mathbb{F}_{2}$.
\end{defn}

\begin{figure}
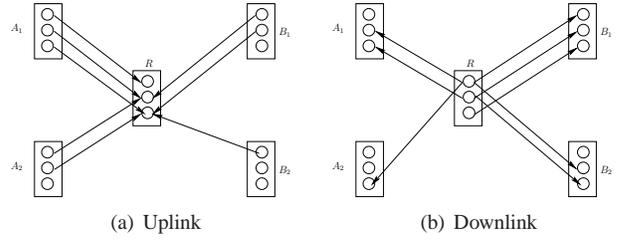

     \centering
     \subfigure[Uplink]{
  \scalebox{0.33}{   \input{detExampleUL.pstex_t}}
}
\subfigure[Downlink]{
       \scalebox{0.33}{\input{detExampleDL.pstex_t}}
}
     \caption{The pictorial representation of a two-pair bidirectional deterministic relay network with channel gains $n_{A_1R}=3$, $n_{B_1R}=2$, $n_{A_2R}=2$, $n_{B_2R}=1$, $n_{RA_1}=2$, $n_{RB_1}=3$, $n_{RA_2}=1$ and $n_{RB_2}=2$. \label{fig:detExample}}
\end{figure}

Now that we have defined the deterministic channel model we can apply it to the multi-pair bidirectional relay network. A pictorial representation of an example of such network with two pairs is shown in Figure \ref{fig:detExample}. In this figure each little circle represents a signal level and what is sent on it is a bit. The transmit and received signal levels are sorted from MSB to LSB from top to bottom. The channel gain between two nodes $i$ and $j$ indicates how many of the first MSB transmitted signal levels of node $i$ are received at destination node $j$. Now as described in the channel model (\ref{eq:channel_model}), at each received signal level, the receiver gets only the modulo two summation of the incoming bits.

\section{Capacity region of the multi-pair bidirectional deterministic relay network}
\label{sec:Main}
\begin{figure}
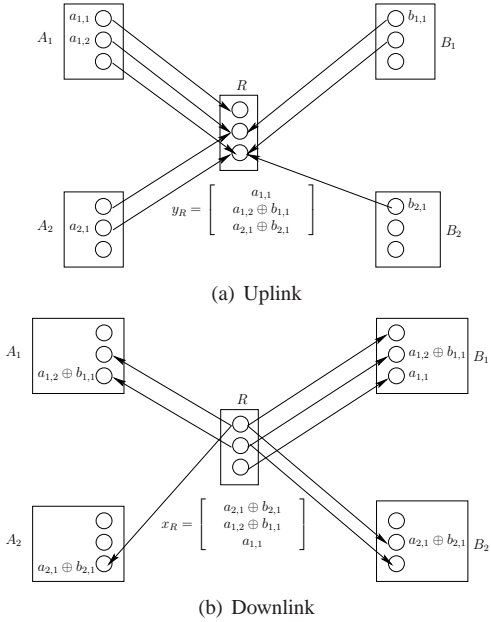

     \centering
     \subfigure[Uplink]{
  \scalebox{0.45}{   \input{detExampleAchiUL.pstex_t}}
}
\hspace{0.2in}
\subfigure[Downlink]{
       \scalebox{0.45}{\input{detExampleAchiDL.pstex_t}}
}
     \caption{The scheme that achieves rate point $(2,1,1,1)$. \label{fig:detExampleAchi}}
\end{figure}

In this section we study the capacity region of the multi-pair bidirectional deterministic relay network. First we state our main result and then in the rest of this section prove it.
\begin{thm}
\label{thm:main}
The capacity region of the multi-pair bidirectional deterministic relay network both (full-duplex and half-duplex), described in Section \ref{sec:sysModel}, is equal to the cut-set upper bound region, and it is achieved by a simple equation forwarding relaying strategy.
\end{thm}
For example in the case that we have only two pairs and the relay is operating on the full-duplex mode, the cut-set upper bound on the capacity region is given by,
{\footnotesize
\begin{align}
R_{A_1B_1} &\leq \min \lp n_{A_1R}, n_{RB_1} \rp  \label{eq: cut-set bound 1} \\
R_{B_1A_1} &\leq \min \lp n_{B_1R}, n_{RA_1} \rp  \label{eq: cut-set bound 2}\\
R_{A_2B_2} &\leq \min \lp n_{A_2R}, n_{RB_2} \rp  \label{eq: cut-set bound 3}\\
R_{B_2A_2} &\leq \min \lp n_{B_2R}, n_{RA_2} \rp  \label{eq: cut-set bound 4}\\
R_{A_1B_1}+R_{A_2B_2} & \leq  \min \lp \max \lp n_{A_1R},n_{A_2R} \rp, \max \lp n_{RB_1},n_{RB_2} \rp\rp  \label{eq: cut-set bound 5}\\
R_{B_1A_1}+R_{B_2A_2} & \leq  \min \lp \max \lp n_{B_1R},n_{B_2R} \rp, \max \lp n_{RA_1},n_{RA_2} \rp\rp  \label{eq: cut-set bound 6}\\
R_{A_1B_1}+R_{B_2A_2} & \leq  \min \lp \max \lp n_{A_1R},n_{B_2R} \rp, \max \lp n_{RB_1},n_{RA_2} \rp\rp  \label{eq: cut-set bound 7}\\
R_{B_1A_1}+R_{A_2B_2} & \leq  \min \lp \max \lp n_{B_1R},n_{A_2R} \rp, \max \lp n_{RA_1},n_{RB_2} \rp\rp  \label{eq: cut-set bound 8}
\end{align}
}
Now consider the network shown in Figure \ref{fig:detExample}. It is easy to check that the rate tuple $(R_{A_1B_1},R_{B_1A_1},R_{A_2B_2},R_{B_2A_2})=(2,1,1,1)$ is inside its cut-set region. In Figure \ref{fig:detExampleAchi} we illustrate a simple scheme that achieves this rate point. With this strategy, the nodes in the uplink transmit
\begin{align*}
&x_{A_1}=\left[ a_{1,1},a_{1,2},0 \right]^t, \quad x_{B_1}=\left[ b_{1,1},0,0 \right]^t \\
&x_{A_2}=\left[ 0,a_{2,1},0 \right]^t ,\quad x_{B_1}=\left[ b_{2,1},0,0 \right]^t
\end{align*}
and the relay receives
\beq y_R=[  a_{1,1} ,~a_{1,2}\oplus b_{1,1},~a_{2,1} \oplus b_{2,1}]^t\eeq
Then the relay will re-order the received equations and transmit
\beq x_R=[a_{2,1} \oplus b_{2,1} ,~a_{1,2}\oplus b_{1,1},~a_{1,1} ]^t\eeq
Then node $A_1$ receives $a_{1,2}\oplus b_{1,1}$ and since it knows $a_{1,2}$ can decode $b_{1,1}$, similarly node $B_1$ can decode $a_{1,1}$ and $a_{1,2}$, node $A_2$ can decode $b_{2,1}$ and finally node $B_2$ can decode $a_{2,1}$. Therefore we achieve the rate point $(2,1,1,1)$.

There are some interesting points about this particular achievability strategy,
\begin{itemize}
\item There is no coding over time,
\item There is no interference between different pairs on the same received signal level at the relay,
\item The relay just re-orders the received equations and forwards them.
\end{itemize}
We call a scheme with these properties an \emph{equation forwarding} scheme. A natural question is whether we can always achieve any rate point in the cut-set bound region with an equation-forwarding scheme?

\begin{figure*}
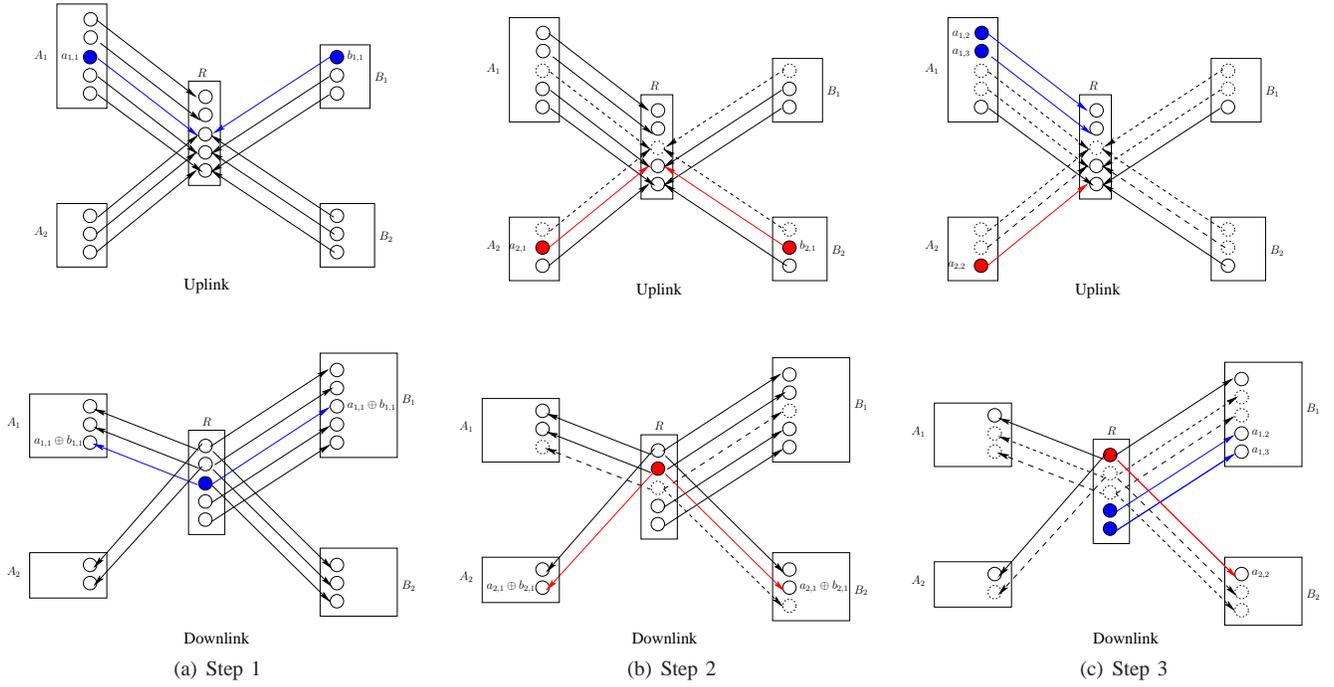

     \centering
     \subfigure[Step 1]{
  \scalebox{0.38}{   \input{detExmInduc1.pstex_t}}
}
\subfigure[Step 2]{
       \scalebox{0.38}{\input{detExmInduc2.pstex_t}}
}
\subfigure[Step 3]{
       \scalebox{0.38}{\input{detExmInduc3.pstex_t}}
}
     \caption{Illustration of the inductive algorithm introduced in Lemma \ref{lem:main} \label{fig:inducExm1}}
\end{figure*}

Quite interestingly, in the next lemma we show that indeed this is always possible.
\begin{lemma}
\label{lem:main}
Any integral\footnote{with integer components.} rate tuple inside the cut-set upper bound on the capacity region of the full-duplex $M$-pair bidirectional deterministic relay network is achieved by a simple equation forwarding scheme.
\end{lemma}
\begin{proof}
Due to lack of space, we only prove the lemma here for $M=2$ pairs, but the proof can be easily generalized to any arbitrary number of pairs. We use induction on the sum rate to show that every integer 4-tuple $(R_{A_1B_1},R_{B_1A_1},R_{A_2B_2},R_{B_2A_2})$ satisfying the cut set bound is achievable by equation forwarding. For convenience we consider two separate cases, and mention how in each case we can assign a relay signal level in uplink and downlink for serving a session worth $1$ bit. We then show that the reduced rate tuple is within the cut-set region of the reduced network. We use an example along the proof to illustrate the ideas. The example network is shown if Figure \ref{fig:inducExm1F}, and one can check that the rate-tuple ($3$,$1$,$2$,$1$) is in its cut-set region.

\emph{Case $1$}: There is a pair where both nodes have nonzero transmission rates. Without loss of generality we may assume that $R_{A_1B_1}$ and $R_{B_1A_1}$ are both nonzero. Our goal is to assign one up-link signal level  and one down-link signal level to the ($A_1$,$B_1$) session.  $A_1$ and $B_1$ will then transmit one bit at the assigned uplink level to the relay, and the relay will transmit the received equation at the assigned down-link level to both $A_1$ and $B_1$.  After doing so and removing the assigned levels, the network will be converted to an equivalent network with lower capacities and reduced rates. If we next show that the new rates satisfy the cut-set bounds of  (\ref{eq: cut-set bound 1}) to (\ref{eq: cut-set bound 8}) for the new network, then by induction we can say that the cut-set bound is achievable in this case.

We claim that in up-link the highest signal level connected to both $A_1$ and $B_1$, and in down-link the lowest signal level connected to both $A_1$ and $B_1$ are the appropriate levels that we are looking for. We refer to these levels by $l_u$ and $l_d$ respectively.  For the sample network of Figure \ref{fig:inducExm1F}  $l_u$ and $l_d$ correspond to the third relay's circle from the bottom in UL,  and third relay's circle from the top in DL respectively. Now the claim is after removing $l_u$ and $l_d$ from UL and DL, the new rate tuple  $(R_{A_1B_1}-1,R_{B_1A_1}-1,R_{A_2B_2},R_{B_2A_2})$ satisfies the cut-set bounds of the new network. First, note that by removing the levels $l_u$ and $l_d$, each of the values $n_{A_1R}$, $n_{B_1R}$, $n_{RA_1}$ and $n_{RB_1}$ will decrease by exactly $1$, and the other four capacities $n_{A_2R}$, $n_{B_2R}$, $n_{RA_2}$ and $n_{RB_2}$ will decrease by at most $1$, depending on the situation. So, after removing the levels and reducing $R_{A_1B_1}$ and $R_{B_1A_1}$ by $1$,  equations (\ref{eq: cut-set bound 1}) and (\ref{eq: cut-set bound 2}) clearly continue to hold . Besides, equations (\ref{eq: cut-set bound 5}) to (\ref{eq: cut-set bound 8}) all remain valid, since in each of them the L.H.S is decremented by $1$ whereas the R.H.S by at most $1$. Because of the symmetry, we only have to resolve the case of equation (\ref{eq: cut-set bound 3}). Let's say (\ref{eq: cut-set bound 3}) is violated because of eliminating the uplink signal level $l_u$. This means prior to this we should have had:  $R_{A_2B_2}\leq n_{A_2R}$ and on the other hand $l_u$ is connected to $A_2$. Because of the choice of $l_u$, the latter means that $n_{A_2R} \geq \min\{n_{A_1R},n_{B_1R}\}$. Let's say $n_{A_2R} \geq n_{A_1R}$.  This and (\ref{eq: cut-set bound 5}) imply that $n_{A_2R} \geq R_{A_1B_1}+R_{A_2B_2}$. But we already have $R_{A_2B_2}=n_{A_2R}$, which means $R_{A_1B_1}=0$, which contradicts our initial assumption. Similarly if $n_{A_2R} \geq n_{B_1R}$, from (\ref{eq: cut-set bound 8}) we conclude $n_{A_2R} \geq R_{B_1A_1}+R_{A_2B_2}$.  This and the fact that  $R_{A_2B_2}=n_{A_2R}$ imply $R_{B_1A_1}=0$, which again cannot be true because of the original assumption.

In a similar way we can show that the removal of $l_d$ can't spoil validity of (\ref{eq: cut-set bound 3}). The reason is if $R_{A_2B_2}=n_{RB_2}$ and $l_d$ is in the transmission range of relay to $B_2$ (i.e. the circle corresponding to level $l_d$ in Figure~\ref{fig:inducExm1F} is connected to $B_2$) then $n_{RB_2} \geq \min\{n_{RA_1},n_{RB_1}\}$. Now if $n_{RB_2} \geq n_{RA_1}$ then from (\ref{eq: cut-set bound 8}) $n_{RB_2} \geq R_{A_2B_2}+R_{B_1A_1}$. But $R_{A_2B_2}=n_{RB_2}$ which means $R_{B_1A_1}=0$. Likewise, if $n_{RB_2} \geq n_{RB_1}$ from equation (\ref{eq: cut-set bound 5}) we conclude $ n_{RB_2} \geq R_{A_1B_1}+R_{A_2B_2}$, which implies $R_{A_1B_1}=0$, which is a contradiction.

Let's apply this scheme to the example network. We should first take the ($A_1$,$B_1$) pair and serve them through one signal level in UL and one level in DL. Next, we take the ($A_2$,$B_2$) and similarly assign corresponding levels in UL and DL to them. These two steps are shown in Figures \ref{fig:inducExm1}a and \ref{fig:inducExm1}b. For clarity, the removed signal levels are dotted in each step.  The remaining unserved rates are ($2$,$0$,$1$,$0$).

\emph{Case 2}: Every session has a node with zero rate. W.L.G., say $R_{B_1A_1}=R_{B_2A_2}=0$ (only $A_1$ and $A_2$ want to communicate with their end parties). We show that $A_1$ and $A_2$ can respectively transmit $R_{A_1B_1}$ and $R_{A_2B_2}$ bits at different signal levels to the relay, and the relay can forward them to $B_1$ and $B_2$ by putting on distinct signal levels in down-link without any interference. In fact, if $n_{A_1R}\leq n_{A_2R}$, then $A_1$ can put $R_{A_1B_1}$ bits on the lowest signal levels in uplink and $A_2$ can put $R_{A_2B_2}$ bits on the levels on top of it. This is feasible because of equations (\ref{eq: cut-set bound 1}) and (\ref{eq: cut-set bound 5}). However, if $n_{A_2R}\leq n_{A_1R}$, $A_2$ can put his bits on the lowest levels and $A_1$ will transmit on top of it. Similarly, in DL if $n_{RB_1} \leq n_{RB_2}$ the bits addressed to $B_1$ will be put on the highest signal levels and the bits addressed to $B_2$ on the signal levels below those of $B_1$. Figure \ref{fig:inducExm1}c shows how this idea is applied to our example network. The final equation forwarding configuration that achieves the rate-tuple for this example is shown in Figure \ref{fig:inducExm1F}.
\end{proof}

Now to prove the main Theorem \ref{thm:main} for the full-duplex scenario, we just need to show that all corner points of the cut-set bound region are achieved by an equation forwarding scheme. Note that since all coefficients of the hyperplanes of the cut-set bound region are integers, then all corner points of the region must be fractional. If a corner point $\overrightarrow{R}$ is integral then by Lemma \ref{lem:main} we know that it is achieved by an equation forwarding scheme. If it is not integral then choose a large enough integer $Q$ such that $Q\overrightarrow{R}$ is integral. Now note that $Q$ instances of a deterministic network overtime, is the same as the original network except all channel gains are multiplied by $Q$. Now since $Q\overrightarrow{R}$ is integral and is obviously inside the cut-set upper bound of the big network (where all channel gains are multiplied by $Q$), then by Lemma \ref{lem:main} it is achievable by an equation-forwarding scheme. This scheme can be simply  translated to an equation forwarding scheme on the original network over $Q$ time-steps. Therefore the corner point $\frac{Q\overrightarrow{R}}{Q}=\overrightarrow{R}$ is achievable.

Similarly we can also prove  Theorem \ref{thm:main} for the half-duplex scenario. In this case relay listens $t$ fraction of the time and transmits the rest. Without loss of generality assume $t$ is a fractional number (otherwise consider the sequence of fractional numbers approaching it). Then choose a large enough integer $Q$ such that $Qt$ is integer. Then consider $Q$ instances of the network over time, such that for $Qt$ instances the relay is listening and in the other $Q(1-t)$ instances it is transmitting. After concatenating these instances together, the resulting network can be thought of as a full-duplex multi-pair network where the uplink channel gains are multiplied by $Qt$ and the downlink channel gains are multiplied by $(1-t)Q$. It is easy to verify that the cut-set bound region of this network is just the cut-set bound region of the original half-duplex network expanded by $Q$. Now by Lemma \ref{lem:main} and the previous argument, we know that the capacity region of this full-duplex multi-pair bidirectional network is equal to its cut-set upper bound region and is achieved by an equation forwarding scheme. Now note that any equation forwarding scheme in this full-duplex network can be translated to an equation forwarding scheme in $Q$ instances of the original half-duplex network; $Qt$ instances the relay is in the listen mode to get the equations and $(1-t)Q$ instances in the transmit mode to send the equations. Therefore the cut-set upper bound region is achievable and the proof is complete.
\section{Remark}
\label{sec:Remark}
\begin{figure}
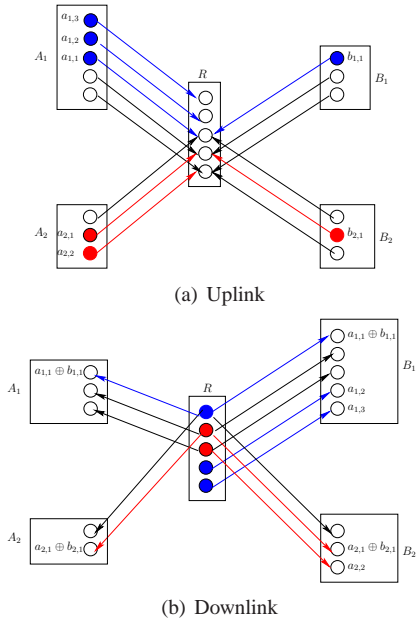

     \centering
     \subfigure[Uplink]{
  \scalebox{0.38}{   \input{detExmInducULF.pstex_t}}
}
\subfigure[Downlink]{
       \scalebox{0.38}{\input{detExmInducDLF.pstex_t}}
}
     \caption{Illustration of the resulting equation forwarding scheme of the inductive algorithm.  \label{fig:inducExm1F}}
\end{figure}

Although the scheme we provided is an inductive way of level assignment and seems unstructured in the sense that it assigns signal levels on a greedy  basis, one can say more about these assignments using certain observations. First of all, note that in this equation forwarding scheme we have in general $2M$ types of equations that the relay might decode. Namely, $M$ types of equations getting bits from one user of a session, and $M$ types of equations getting bits from both users of the same pair. Refer to the example network of Figures \ref{fig:inducExm1} and \ref{fig:inducExm1F}, and observe that in the final configuration all equations of the same type were concatenated together both in UL and DL. In general one can serve all equations of the same type at once by choosing a pair with nonzero rates and serving them until one of the rates is zero. For example, assuming $R_{A_1B_1}\leq R_{B_1A_1}$ and $R_{A_2B_1}\leq R_{B_2A_2}$, instead of reducing ($R_{A_1B_1}$,$R_{B_1A_1}$,$R_{A_2B_2}$,$R_{B_2A_2}$) to ($R_{A_1B_1}-1$,$R_{B_1A_1}-1$,$R_{A_2B_2}$,$R_{B_2A_2}$) one can reduce it to (0,$R_{B_1A_1}-R_{A_1B_1}$,$R_{A_2B_2}$,$R_{B_2A_2}$) all at once. Then the same thing can be done for the other pair. Now by rearranging the signal levels, it is easy to show that in the final configuration, all the equations of the same type are in concatenation. The conclusion of the above argument is that for any network configuration, one can always find $2M$ groups of disjoint signal levels (both in UL and DL) with the following properties:

\begin{center} $\vdots$
\end{center}
\begin{itemize}
\item Group $2i-1$: $\min\{R_{A_{i}B_{i}},R_{B_iA_i}\}$ levels connected to both $A_{i}$ and $B_{i}$.
\item Group $2i$: $\max\{R_{A_{i}B_{i}},R_{B_iA_i}\}-\min\{R_{A_{i}B_{i}},R_{B_iA_i}\}$ levels connected to the one of $A_i$ and $B_i$ with higher transmission rate.
\end{itemize}
\begin{center}
$\vdots$
\end{center}

Furthermore, levels of each group are all in concatenation. Intuitively this observation suggests that in the noisy case each user need to only break the transmit signal to at most 2M different levels, independent of the channel gain strengths.

\section{Conclusions}
In this paper we studied the multi-pair bidirectional relay network which is a generalization of the bidirectional relay channel. We examined this problem in the context of the deterministic channel model introduced in \cite{ADT072} and characterized its capacity region completely in both full-duplex and half-duplex cases. We also showed that the capacity can be achieved by a simple equation-forwarding strategy and illustrated some structures on the signal levels that these equations are created at. In ongoing work, we have made progress on using these insights to find an approximate capacity characterization of the noisy (Gaussian) version of this problem. We hope to completely answer this question in a future work.
\label{sec:Conc}

{
\bibliographystyle{IEEE}

}

\end{document}